\documentclass[twocolumn,twoside,slac_two]{revtex4}
\usepackage{graphicx}
\usepackage{fancyhdr}
\begin{document}
\newcommand{\nc}{\newcommand}
\nc {\beq}{\begin{equation}} \nc {\eeq}{\end{equation}} \nc
{\ra}{\rightarrow} \nc{\ch}{\rm H^{\pm}} \nc{\tnb}{\rm tan\beta}
\nc{\ttbar}{\rm t\bar{t}} \nc{\wj}{\rm W^{\pm}+3~jets}

\title{Search for the Light Charged Higgs in CMS}

%

\author{M. Hashemi}
\affiliation{IPM, P.O.Box 19395-5531, Tehran, Iran}

\begin{abstract}
In this report the CMS potential for the light charged Higgs boson
discovery in the Minimal Supersymmetric Standard Model is presented.
First the latest results of the Tevatron and LEP experiments on the
light charged Higgs search are reminded. In the rest of the report
the perspectives of CMS for the light charged Higgs search are
presented with description of some details of the analysis. The
results are based on the full simulation and reconstruction of the
CMS detector including the systematic uncertainties on the
background determination. Finally the $\rm 5\sigma$ discovery
contour for an integrated luminosity of $\rm 30fb^{-1}$ is shown.
\end{abstract}

\maketitle

\thispagestyle{fancy}

\section{Introduction}
The Minimal Supersymmetric extension to the Standard Model
(MSSM)~\cite{MSSM} requires the introduction of two Higgs doublets
in order to preserve the supersymmetry. Five physical Higgs
particles are predicted, two CP-even (h,H), one CP-odd (A) and two
charged ones ($\rm H^{\pm}$). All couplings and masses of the MSSM
Higgs sector are determined at the lowest order by two independent
parameters, which are generally chosen as $\rm tan\beta = \upsilon_2
/ \upsilon_1$, the ratio of the vacuum expectation values of the two
Higgs doublets, and the pseudo--scalar Higgs boson mass $m_{\rm A}$.
Since the charged Higgs boson is a crucial signature of the MSSM, it
is presently considered as one of the most interesting particles in
the MSSM Higgs sector.

\begin{figure}[h]
\centering
  \resizebox{5cm}{5cm}{\includegraphics{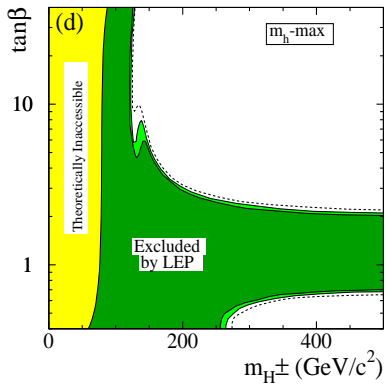}}
  \caption{The excluded region at 99.7$\%$ C.L. (Dark Green) obtained by the LEP Higgs working group.}
  \label{LEP}
\end{figure}

\begin{figure}[h]
\centering
  \resizebox{5cm}{5cm}{\includegraphics{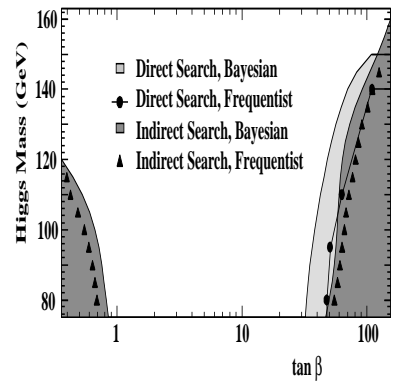}}
  \caption{The excluded region at 95$\%$ C.L. obtained by D0 and CDF at Tevatron.}
  \label{TEV}
\end{figure}

The latest results from the LEP Higgs working group concerning the
indirect search for neutral MSSM Higgs bosons provide a lower mass
limit of 125~GeV/$c^2$ for $m_{\ch}$ at $\rm \tnb > 10$~\cite{lep}.
Results from direct searches at LEP give instead a lower limit on
the charged Higgs boson mass of $\sim 90~$GeV/$c^{2}$ for $\rm
BR(H^{+}\ra \tau \nu) \simeq 1$ \cite{directlep}. Figure \ref{LEP}
shows the excluded region obtained in 2006 by the LEP Higgs working
group.
Direct and indirect searches performed at the Tevatron lead to some
exclusion region in the parameter space for $\rm \tnb > 50 $ and
$\rm \tnb<1$ \cite{d0,cdf}. Figure \ref{TEV} shows the excluded
region obtained by Tevatron in the D0 and CDF experiments.
In the study presented in this report the light charged Higgs $\rm
(m_{\ch}<m_t-m_b)$ is searched for in the channel ${\rm H}_{\rm
MSSM}^{\pm}\rightarrow \tau^{+}\nu_{\tau}$, $\tau\rightarrow$hadrons
in the mass range $125<m_{\rm H^{\pm}}<170~$ GeV/$c^2$
\cite{mynote}. The mass range chosen is the most critical since for
Higgs boson masses close to the top mass the signal rate is low.
Proving that the discovery is attainable in this mass range would
imply that the light charged Higgs boson can be discovered also at
lower masses. All results are obtained according to the LEP $m_{\rm
h^0}-{\rm max}$ benchmark scenario \cite{lep} with the following
choice of the MSSM parameters: SU(2) gaugino mass
$M_2=200~$GeV/$c^2$, $\rm \mu=+200$~GeV/$c^2$, gluino mass
$M_{\tilde{g}}=800~$GeV/$c^2$, SUSY breaking mass parameter $M_{\rm
SUSY}= 1~$TeV/$c^2$ and stop mixing parameter $X_{\rm
t}=\sqrt{6}~M_{\rm SUSY}~ (X_{\rm t}=A_{\rm t}+\mu cot\beta)$. The
$m_{\rm h^0}-{\rm max}$ scenario is one of the LEP benchmark
scenarios designed to maximize the upper limit on $m_{\rm h^{0}}$ by
non-zero stop mixing parameter thus providing a wider available
parameter space compared to the case of $no-mixing$ scenario in
which the stop mixing parameter is set to zero and the available
parameter space is smaller. The top quark mass is set to
175~GeV/$c^2$.

\section{Signal and background simulation, cross sections and branching ratios}
\begin{figure*}[t]
\centering
  \resizebox{10cm}{5cm}{\includegraphics{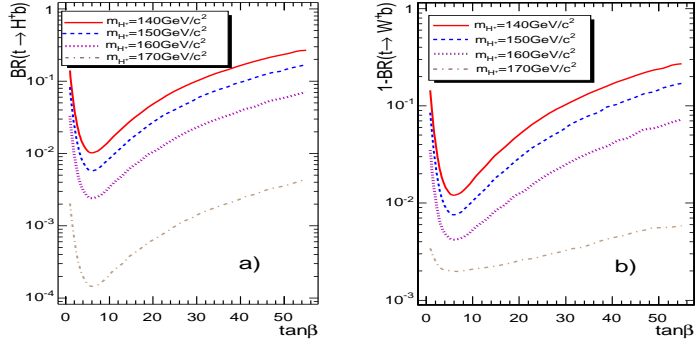}}
  \caption{a)Branching ratio of top decay to $\rm \ch$ vs $\rm \tnb$, b)Branching ratio of top decay to $\rm W^{\pm}$ vs $\rm \tnb$}
  \label{BRt2Hb}
\end{figure*}

\begin{figure}[h]
\begin{center}
    \resizebox{5cm}{5cm}{\includegraphics{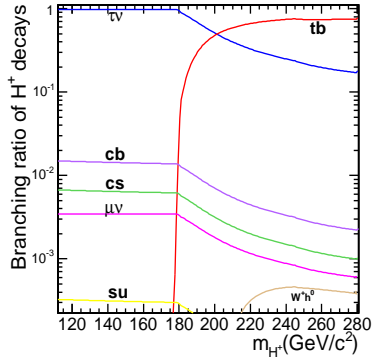}}
  \caption{Branching ratio of $\rm \ch$ decays for different $m_{\ch}$ at $\rm tan \beta =20$}
  \label{htaunu}
\end{center}
\end{figure}

The MSSM charged Higgs boson can be produced in top quark decays,
$\rm t\ra H^{+} b$, if $m_{\ch}<m_{\rm t}-m_{\rm b}$. In this report
the case of $\rm W^{\pm}$ leptonic decay (e or $\mu$) is described.
The background channels consist of $\rm \ttbar$ events with at least
a single lepton (e or $\rm \mu$) and $\rm \tau$-jets or jets which
could fake $\rm \tau$-jets, $\wj$ events and also single top events
which have, however, a small contribution. A total NLO $\rm t
\bar{t}$ cross section of $\rm 850~pb$ is used in this
report~\cite{ttbar} for calculating the cross sections of the
different final states. The cross section of $\wj$ events is
obtained with the MadEvent generator~\cite{madevent} with some
preselection cuts: $E^{\rm jet}_T> 20~{\rm GeV}, |\eta| <2.5$ for
all final state jets and $R=\sqrt{(\Delta \eta)^2 + (\Delta \phi)^2}
= 0.5$ for all pairs of final state jets. The branching ratio of top
decay to charged Higgs depends on both $m_{\ch}$ and $\tan\beta$ as
shown in Fig. \ref{BRt2Hb}a. The corresponding top decay to $\rm
W^{\pm}b$ (Fig. \ref{BRt2Hb}b) decreases with increasing $\rm
tan\beta$ while keeping the sum of the branching ratios around one
(for $m_{\rm H^{+}}\simeq 170~$GeV/$c^2$ the  $\rm t\ra W^{+}s$
branching ratio is of the same order as that of $\rm t\ra H^{+} b$
therefore these two plots are not similar for $m_{\rm H^{+}}=
170~$GeV/$c^2$).
Figure \ref{htaunu} shows the $\rm \ch$ decay branching ratio in
different final states for $\rm tan\beta = 20$.
\begin{table}[hbt!]
\caption{Cross section times branching ratio of $\rm t \bar{t}\ra
H^{\pm}W^{\mp}b \bar{b}\rightarrow \tau \nu_{\tau}\ell \nu_{\ell} b
\bar{b}, \tau \ra hadrons$ for $\rm \tnb =20$} \label{tbl1}
\begin{center}
\begin{tabular}{|c|c|c|c|c|}
\hline $m_{\ch}($GeV/$c^2)$ & $140$ & $150$ & $160$
& $170$ \\
\hline
Cross section [pb] & 10.70 & 5.06 & 1.83 & 0.157\\
\hline
\end{tabular}
\end{center}
\end{table}

\begin{table}[hbt!]
\caption{Cross section times branching ratio of $\rm t \bar{t}\ra
H^{+}H^{-}b \bar{b} \rightarrow \tau^+ \nu_{\tau}\tau^-
\bar{\nu}_{\tau} b \bar{b}, \tau^{\pm} \ra hadrons~(leptons)$ for
$\rm \tnb =20$} \label{tbl2}
\begin{center}
\begin{tabular}{|c|c|c|c|}
\hline
$m_{\ch}($GeV/$c^2)$ & 140 & 150 & 160 \\
\hline
Cross section (pb) & 0.91 & 0.19 & 0.02 \\
\hline
\end{tabular}
\end{center}
\end{table}

Table~\ref{tbl1} shows the cross section times branching ratio of
$\rm t\bar{t}\ra \ch W^{\mp} b \bar{b}$ events for $\rm \tnb=20$.
Due to the small cross section of $\rm t\bar{t}\ra
H^{+}H^{-}b\bar{b}$ as shown in Table \ref{tbl2} only $\rm
t\bar{t}\ra \ch W^{\mp} b \bar{b}$ events were generated and
simulated as signal events for $m_{\ch}<170~$GeV/$c^2$. For
$m_{\ch}=170~$GeV/$c^2$ both $\rm t\bar{t} + gb$ and $\rm gg \ra t
\bar{b} H^{\pm}$ are used for comparison. The NLO cross section
times branching ratio of signal events with $m_{\rm H^{+}}\simeq
m_{\rm t}$ is listed in Table~\ref{tbl3}. Finally the cross section
of the main background channels are shown in Table~\ref{tbl4}.

\begin{table*}[t]
\caption{Cross section times branching ratio of signal events for
$m_{\ch} \simeq m_{\rm t}$ according to NLO calculations
in~\cite{tilman} for $\rm \tnb=20$.} \label{tbl3}
\begin{center}
\begin{tabular}{|c|c|c|}
\hline Channel & \begin{minipage}{4cm}
\centering $\rm gb\ra t H^{\pm} \rightarrow \ell \nu_{\ell}b\tau\nu_{\tau}$\\
\centering ($\rm \tau \ra hadrons$)\\
\centering $m_{\ch}=170$~GeV/$c^2$
\end{minipage}
& \begin{minipage}{4cm}
\centering $\rm gg \ra t \bar{b} H^{\pm} \rightarrow \ell \nu_{\ell}b\bar{b}\tau\nu_{\tau}$\\
\centering ($\rm \tau \ra hadrons$)\\
\centering $m_{\ch}=170$~GeV/$c^2$
\end{minipage} \\
\hline
Cross section [pb] & 0.14 & 0.297  \\
\hline
\end{tabular}
\end{center}
\end{table*}

\begin{table*}[t]
\caption{Cross section times branching ratio of background events}
\label{tbl4}
\begin{center}
\begin{tabular}{|c|c|c|c|c|}
\hline Channel & \begin{minipage}{2.5cm}\centering $\rm t \bar{t}\ra
W^{+}W^{-}b \bar{b}$\\ \centering $\rm \ra \ell \nu_{\ell} \tau
\nu_{\tau} b \bar{b}$\\ \centering ($\rm \tau \ra hadrons$)
\end{minipage} & \begin{minipage}{2.5cm}\centering $\rm t \bar{t}\ra
W^{+}W^{-}b \bar{b}$\\ \centering $\rm \ra \ell \nu_{\ell} \ell'
\nu_{\ell'} b \bar{b}$\\ \centering $\rm \ell,\ell' = e ~ or~ \mu$
\end{minipage} & \begin{minipage}{2.5cm}\centering $\rm t \bar{t}\ra
W^{+}W^{-}b \bar{b}$\\ \centering $\rm \ra \ell \nu_{\ell} jj b
\bar{b}$ \end{minipage}
& \begin{minipage}{2.5cm}\centering $\rm W^{\pm}+3~jets$ \\ \centering $\rm W^{\pm}\ra e~or~\mu$\end{minipage}\\
\hline
Cross section [pb] & 25.8 & 39.7 & 245.6 & 840 \\
\hline
\end{tabular}
\end{center}
\end{table*}
\section{Event selection}
Full simulation of the detector response was achieved with CMSIM
\cite{cmsim} or OSCAR \cite{oscar} including pile--up events
corresponding to a luminosity of $\rm 2 \times 10^{33}
cm^{-2}s^{-1}$. The full reconstruction of the detector response and
event analysis was performed with the $\rm ORCA$
package~\cite{orca}.
The event selection is as the following:\\
\begin{itemize}
\item Event trigger\\
Events are triggered if there is a muon with $p_{T}>14$~GeV,
$|\eta|<2.1$ or an electron with $p_{T}>29$GeV, $|\eta|<2.4$. At HLT
harder cuts are applied such as $p_{T}>19$~GeV and isolation
requirements for muons \cite{daq}.
\item Jet reconstruction, selection and b-tagging\\
Jets were reconstructed with the Iterative Cone algorithm~\cite{daq}
with a cone size $R=\sqrt{(\Delta \eta)^2 + (\Delta \phi)^2} = 0.5$.
Jets were selected within $|\eta| <2.4$ and requiring $E^{\rm jet}_T > 40~{\rm GeV}$. There should be at least three jets in the event to be accepted.  \\
The b-tagging is done by Impact Parameter Significance method and
one b-jet is required in the event with kinematic cuts the same as
those applied on jets.
\item Offline $\rm \tau$ tagging\\
The $\tau$ reconstruction and selection is performed offline
starting from the Level 1 $\rm \tau$-like energy deposits in the
calorimeters (\cite{daq}). A regional jet reconstruction is
performed around the direction of the Level 1  $\rm \tau$ objects;
the raw jet tranverse energy is required to be  $E^{\rm
jet}_T>20~{\rm GeV}.$ The contribution from electrons faking
$\tau$-jets can be reduced by requiring the hottest HCAL tower $\rm
E_T$ to be larger than $\rm 2~GeV$ (only towers belonging to jets
are considered). In the next step a matching cone with $\rm \Delta
R=0.1$ is defined around the jet axis. Tracks with $p_T>1$~GeV/$c$
are searched for within the matching cone and the largest $p_T$
track is identified and a signal cone with $\rm \Delta R=0.07$ is
defined around its direction. The signal vertex is considered as the
vertex from which the leading track originates. Then an isolation
cone with $\rm \Delta R=0.4$ is defined around the leading track to
check the tracker isolation requirement. If no track from the signal
vertex is found with $p_T>1$~GeV/$c$ in the isolation cone except
for those falling in the signal cone, the isolation is satisfactory.
The electromagnetic isolation parameter (Eq. \ref{pisol}) is used to
suppress the contamination arising from quark and gluon
jets~\cite{Sasha}:
\begin{equation}
\label{pisol}
\begin{tabular}{l}
$P_{\rm isol.} = $ \\
$ \displaystyle \sum_{{\rm crystals}, \Delta
R_{{\rm crystal},\tau-{\rm jet}}<0.4}{{E_T}_{\rm crystal}}-$ \\
$\displaystyle  \sum_{{\rm crystals}, \Delta R_{{\rm
crystal},\tau-{\rm jet}}<0.13}{{E_T}_{\rm crystal}}<5.6~{\rm GeV}$
\end{tabular}
\end{equation}

When a jet satisfies all requirements mentioned above, it is
identified as a $\rm \tau$-jet candidate. A cut on the transverse
energy of the $\rm \tau$ jet, ${E_{T}}_{\tau}>40$~GeV, can be
applied to suppress the background. Since the leading track in the
jet is more energetic in signal events than in background events, a
cut on the $\rm \tau$ energy carried by the leading track in the
signal cone is applied by requiring $p_{\rm leading ~
track}/E_{\tau}>0.8$. The total charge of all tracks contained in
the signal cone of an identified  $\rm \tau$-jet gives an estimate
of the $\rm \tau$-lepton charge. The sum of the charges of the event
trigger lepton  (${\rm e\;or \mu}$) and of the $\rm \tau$-lepton is
hence required to be $Q(\ell)+Q(\tau)=0$, since they originate from
mother particles of opposite charges ($\rm H^{\pm},W^{\mp}$).
\item Missing ${\boldmath E_T}$ reconstruction\\
The missing transverse energy reconstruction is performed by summing
all the Ecal energy deposits plus the Hcal towers and correcting for
muons (if any in the event). The raw energy of jets involved in the
$E^{\rm miss}_T$ reconstruction are corrected~\cite{type1} to
improve their energy resolution. Only jets with raw jet $E_T>30$~GeV
are used; This cut is optimized to improve the $E^{\rm miss}_T$
resolution. A cut at $E_{T}^{\rm miss}>70$~GeV is applied to
increase the signal to background ratio.
\end{itemize}
\section{Signal and background selection efficiencies}
The summary of all event selection requirements together with the
corresponding efficiencies is given in Tables \ref{seleff1},
\ref{seleff2} (signal) and Table \ref{seleff3} (background). As
mentioned in Section 2 the number of signal events  from $\rm
t\bar{t}$ and gb fusion for $m_{\ch}=170$~GeV/$c^2$ should actually
be added together and then compared to $\rm gg$ fusion process. No
significant difference is seen between  $\rm t\bar{t} +$gb and gg.
The one b-jet requirement mentioned in Section 3.2 is decomposed
into two different steps: firstly events with at least one b-jet are
accepted, secondly events with more than one b-jet are rejected. The
first requirement has a higher selection efficiency for background
events, however, the second cut rejects background events reasonably
(more than $\rm \sim30\%$) at the expense of losing $\rm \sim
5-20\%$ of signal events. The efficiency of the second cut for
signal events increases with larger $m_{H^{\pm}}$, i.e. with lower
signal cross section. Hence there is little loss where the signal
rate is low.

Results can be extrapolated to other $\rm \tnb$ values by using the
branching ratios of Figs. \ref{BRt2Hb}a, \ref{BRt2Hb}b and
\ref{htaunu}. The effect of a larger $\rm \tnb$ appears only as some
increase in the signal rate and no kinematic effect is expected.
Selection efficiencies are then expected to be independent from $\rm
\tnb$. Different $\rm \tnb$ values were investigated in this study
for different masses of charged Higgs. In the next sections the
results are presented and the final conclusions are drawn.
\begin{table*}[t]
\begin{center}
\caption{List of selection cuts and their efficiencies for signal
events with $m_{\ch}<170$~GeV/$c^2$ for $\rm \tnb=20$. Numbers in
each row show the remaining cross section after applying the
corresponding cut. Numbers in parentheses are relative efficiencies
in percent.} \label{seleff1}
\begin{tabular}{|c|c|c|c|}
\hline $ $ & $\rm t \bar{t}\ra H^{\pm}W^{\mp}b \bar{b}$ & $\rm t
\bar{t}\ra H^{\pm}W^{\mp}b \bar{b}$ & $\rm t \bar{t}\ra
H^{\pm}W^{\mp}b \bar{b}$
\\
$ $& $\rm ~\ra \ell \nu_{\ell} \tau \nu_{\tau} b \bar{b}$&$\rm ~\ra
\ell \nu_{\ell} \tau \nu_{\tau} b \bar{b}$&$\rm ~\ra \ell \nu_{\ell}
\tau \nu_{\tau} b \bar{b}$
\\
$ $
& $ m_{\ch}=140$~GeV/$c^2$ &$  m_{\ch}=150$~GeV/$c^2$&$ m_{\ch}=160$~GeV/$c^2$\\
\hline
$\rm \sigma \times BR[fb]$ & 10.7 $\rm \times 10^3$ & 5060. & 1830. \\
\hline
L1 + HLT & 5170.5(48.3) & 2456.3(48.5) & 888.9(48.6)\\
\hline
$\rm >=3$ jets & 1889.7(36.5) & 795.0(32.4) & 264.3(29.7)\\
\hline
$\rm \ge 1~b~jet$ & 1103.5(58.4) & 427.4(53.8) & 131.4(49.7) \\
\hline
$\rm < 2~b~jets$ & 883.0(80.0) & 358.7(83.9) & 119.2(90.7) \\
\hline
Having L1 $\rm \tau$ & 878.4(99.5)& 357.4(99.6)& 119.0(99.8) \\
\hline
$\rm \tau$-jet reconstruction & 875.0(99.6) & 356.5(99.7) & 118.8(99.8)\\
\hline
Hottest HCAL tower $E_T > 2 $~GeV & 778.0(88.9)& 316.1(88.6) & 105.9(89.1) \\
\hline
Tracker isolation & 378.2(48.6) & 163.5(51.7)& 52.7(49.8)\\
\hline
Ecal isolation & 292.9(77.4) & 134.2(82.1) & 43.1(81.8)\\
\hline
$\tau~~E_T > 40$~GeV & 244.3(83.4) & 113.0(84.2) & 36.5(84.7)\\
\hline
$p_{\rm leading~track}/E_{\tau}>0.8$ & 102.3(41.9)& 50.7(44.8) & 16.8(45.9)\\
\hline
$ Q(\ell) + Q(\tau)=0 $ & 88.0(86.0) & 42.4(83.6) & 14.6(87.0)\\
\hline
$   E_{T}^{\rm miss}>70$~GeV  & 51.0(58.0) & 25.4(59.9) & 9.2(63.3)\\
\hline
\begin{minipage}{3.cm} Expected Number of events after $\rm 10~fb^{-1}$\end{minipage} & 510 & 254 & 92  \\
\hline
\end{tabular}
\end{center}
\end{table*}

\begin{table*}[t]
\caption{List of selection cuts and their efficiencies for signal
events with $m_{\ch}=170$~GeV/$c^2$ for $\rm \tnb=20$. Numbers in
each row show the remaining cross section after applying the
corresponding cut. Numbers in parentheses are relative efficiencies
in percent.} \label{seleff2}
\begin{center}
\begin{tabular}{|c|c|c|c|}

\hline $ $ & $\rm t \bar{t}\ra H^{\pm}W^{\mp}b \bar{b}$ & $\rm gb
\ra tH^{\pm}$ & $\rm gg \ra t\bar{b}H^{\pm}$
\\
$ $&$\rm ~\ra \ell \nu_{\ell} \tau \nu_{\tau} b \bar{b}$&$\rm ~\ra
\ell \nu_{\ell} \tau \nu_{\tau} b$&$\rm ~\ra \ell \nu_{\ell} \tau
\nu_{\tau} b \bar{b}$
\\
$ $
&$  m_{\ch}=170$~GeV/$c^2$&$  m_{\ch}=170$~GeV/$c^2$&$ m_{\ch}=170$~GeV/$c^2$\\
\hline
$\rm \sigma \times BR[fb]$ &  157. & 140. & 297. \\
\hline
L1 + HLT & 78.0(49.7)& 70.5(50.4)& 145.4(48.9)\\
\hline
$\rm >=3$ jets & 23.2(29.7)& 21.7(30.7)& 55.3(38.0)\\
\hline
$\rm \ge 1~b~jet$ & 11.5(49.4) & 11.7(54.1) & 31.9(57.7) \\
\hline
$\rm < 2~b~jets$ & 10.9(94.8) & 10.0(85.5) & 25.8(80.9) \\
\hline
Having L1 $\rm \tau$ & 10.8(99.8)& 10.0(99.6)& 25.7(99.4)\\
\hline
$\rm \tau$-jet reconstruction & 10.8(99.9)& 10.0(99.9)& 25.5(99.1)\\
\hline
Hottest HCAL tower $E_T > 2$GeV& 9.6(88.4) & 8.9(88.8) & 22.6(88.9)\\
\hline
Tracker isolation & 4.9(51.3)& 5.1(57.2)& 11.4(50.5)\\
\hline
Ecal isolation & 4.2(84.9)& 4.3(84.5)& 9.6(84.4)\\
\hline
$\tau~~E_T > 40$GeV & 3.8(90.9)& 3.9(90.6)& 8.6(89.2)\\
\hline
$p_{\rm leading~track}/E_{\tau}>0.8$ & 1.6(41.7)& 1.8(45.9)& 3.4(39.6)\\
\hline
$Q(\ell) + Q(\tau)=0 $ & 1.3(84.4)& 1.6(87.2)& 2.8(82.6)\\
\hline
$ E_{T}^{\rm miss}>70$~GeV & 0.8(61.7)& 1.0(65.2)& 1.6(55.3)\\
\hline
\begin{minipage}{3.cm} Expected Number of events after $\rm 10~fb^{-1}$\end{minipage} & 8 & 10 & 16 \\
\hline

\end{tabular}
\end{center}
\end{table*}
\begin{table*}[h]
\caption{List of selection cuts and their efficiencies for
background events. Numbers in each row show the remaining cross
section after applying the corresponding cut. N umbers in
parentheses are relative efficiencies in percent.} \label{seleff3}
\begin{center}
\begin{tabular}{|c|c|c|c|c|}

\hline $ $ & $\rm t \bar{t}\ra W^{+}W^{-}b \bar{b}$ & $\rm t
\bar{t}\ra W^{+}W^{-}b \bar{b}$ & $\rm t \bar{t}\ra W^{+}W^{-}b
\bar{b}$ & $\rm \wj$
\\
$ $& $\rm ~\ra \ell \nu_{\ell} \tau \nu_{\tau} b \bar{b}$&$\rm ~\ra \ell \nu_{\ell} \ell' \nu_{\ell'} b \bar{b}$&$\rm ~\ra \ell \nu_{\ell} jj b \bar{b}$&$\rm W^{\pm} \ra \ell \nu_{\ell}$
\\
\hline
$\rm \sigma \times BR[fb]$ & 25.8 $\rm \times 10^3$ & 39.8 $\rm \times 10^3$ & 245.6$\rm \times 10^3$ & 840.$\rm \times 10^3$\\
\hline
L1 + HLT & 12101.2(46.9)& 28429.1(71.4)&99506.6(40.5)& 287280(34.2)\\
\hline
$\rm >=3$ jets & 5105.2(42.2)& 11306.6(39.8)& 66038.6(66.4)&114050(39.7)\\
\hline
$\rm \ge 1~b~jet$ & 3428.3(67.1) & 7622.0(67.4)& 43433.0(65.8)& 24292.7(21.3) \\
\hline
$\rm < 2~b~jets$ & 2325.7(67.8)& 5262.7(69.0)& 29003.4(66.8)& 21207.5(87.3) \\
\hline
Having L1 $\rm \tau$ & 2310.7(99.3)& 5233.7(99.4)&28698.8(98.9)& 20613.7(97.2)\\
\hline
$\rm \tau$-jet reconstruction & 2303.6(99.7)&5224.4(99.8)&28465.0(99.2)&19438.7(94.3)\\
\hline
\small{Hottest HCAL tower $E_T > 2$GeV } & 2034.1(88.3)&3850.6(73.7)&26635.1(93.6)&17125.5(88.1)\\
\hline
Tracker isolation &798.7(39.3)&1120.6(29.1)&6653.3(25.0)& 5411.7(31.6)\\
\hline
Ecal isolation & 545.6(68.3)&519.5(46.3)&2952.8(44.4)& 2554.3(47.2)\\
\hline
$\tau~~E_T > 40$GeV& 405.8(74.4)&341.8(65.8)&1946.8(65.9)& 1312.9(51.4)\\
\hline
$p_{\rm leading~track}/E_{\tau}>0.8$ & 123.5(30.4)&131.9(38.6)&377.9(19.4)& 224.5(17.1)\\
\hline
$Q(\ell) + Q(\tau)=0 $ & 95.7(77.5)&56.7(43.0)&78.8(20.9)& 27.1(12.1)\\
\hline
$ E_{T}^{\rm miss}>70$~GeV  & 51.6(53.9)& 29.3(51.8)&36.6(46.4)& 10.7(39.3)\\
\hline
\begin{minipage}{3.cm} Expected Number of events after $\rm 10~fb^{-1}$\end{minipage} & 516 & 293 & 366 & 107 \\

\hline
\end{tabular}
\end{center}
\end{table*}

\section{Contribution of other signal and background final states}
Contribution of other signal final states is estimated to be an
additional factor of $\sim 2\%$ to the number of signal events
\cite{mynote}. The single top background produces about 60 events
after all selection cuts and is taken into account in all
calculations. The contribution of other background events such as
$\rm W^{\pm}bb$ with $\rm W^{\pm} \ra e\nu_{e}~ or~ \mu\nu_{\mu}$
and $\rm Zbb$ followed by the decay $\rm Z\ra ee~ or~ \tau\tau$ are
also estimated to be negligible.
\section{Results with systematic uncertainties}
Since the number of background events after all selection cuts is
large enough (more than 1000 events), the significance can be
defined as
\begin{equation}
 S=\frac{{N^{\rm MSSM}_{\rm obs.}-(N^{\rm SM}_{\rm W+3~jets}+N^{\rm SM}_{\rm t\bar{t}})}}{\sqrt{N^{\rm SM}_{\rm W+3~jets}+N^{\rm SM}_{\rm t\bar{t}}+{(\Delta N^{\rm SM}_{\rm W+3~jets})}^2+{(\Delta N^{\rm SM}_{\rm t\bar{t}})}^2}}
\label{sigsys}
\end{equation}
The approach for including systematic uncertainties in the signal statistical significance calculation have been described in detail in \cite{mynote}. Therefore here only the sources of uncertainties and how they are taken into account are presented.\\
The total systematic uncertainty on the  $\rm t\bar{t}$ background
can be expressed as
\begin{equation}
\label{ttunc}
\begin{tabular}{l}
$\Delta_{\rm sys.}^{\ttbar}  =\Delta_{\rm lepton~ reconstruction}
\oplus \Delta_{\geq \rm 3~jet~selection} $\\
$ \oplus \Delta_{\rm 1 ~b-jet~ tagging} \oplus \Delta_{\rm 1 ~\tau~
tagging}\oplus \Delta_{\rm lumi.} \oplus \Delta^{\ttbar}_{\rm
theo.}$
\end{tabular}
\end{equation}
The $\wj$ background uncertainty is estimated as the following
\begin{equation}
\label{wjuncertainty}
\begin{tabular}{l}
$ \Delta_{\rm sys.}^{\wj}  =\Delta_{\rm stat.} \oplus \frac{\Delta
N^{\ttbar}_{B}}{N^{\wj}_{B}} \oplus \Delta_{\rm 3~non-b-jet} $ \\
$\oplus \Delta_{\rm b-jet~ mistagging} \oplus \Delta_{\tau ~ \rm
mistagging}$
\end{tabular}
\end{equation}
In Eq. \ref{wjuncertainty}, $\Delta N^{\ttbar}_{B}$ is the
uncertainty of the number of $\ttbar$ events in the
background area and $N^{\wj}_{B}$ is the number of $\wj$ events in the background area.\\
Table \ref{uncertainties} shows the values of individual
uncertainties which were used in the signal significance calculation
for $\rm 30~fb^{-1}$. These are the minimum values of uncertainties
in the duration of an integrated luminosity of $\rm 30~fb^{-1}$
which would be reached at the end of this period.
\begin{table*}[t]
\caption{The values of different experimental and theoretical
uncertainties used for $\ttbar$ and $\wj$ background events at $\rm
30~fb^{-1}$.} \label{uncertainties}
\begin{center}
\begin{tabular}{|l|c|}
\hline
Scale uncertainty of $\ttbar $ cross section & $\rm 5\%$ \\
\hline
PDF uncertainty of  $\ttbar $ cross section & $\rm 2.5\%$ \\
\hline
b-tagging & $\rm 5\%$ \\
\hline
$\rm \tau$-tagging & $\rm 4\%$ \\
\hline
Lepton identification & $\rm 2\%$ \\
\hline
Jet energy scale & $\rm 3\%$ \\
\hline
Mistagging a non-b-jet as a b-jet & $\rm 5\%$ \\
\hline
Mistagging a jet as a $\rm \tau$-jet & $\rm 2\%$ \\
\hline
Non-b-jet identification (anti-b-tagging) & $\rm 5\%$ \\
\hline
Luminosity uncertainty & $\rm 5\%$ \\
\hline
\end{tabular}
\end{center}
\end{table*}

\subsection{Final result and the ${\boldmath 5 \sigma}$ discovery contour with systematic uncertainties}
Equation~\ref{sigsys} is used to calculate the signal statistical
significance at $\rm 30fb^{-1}$ for different $\rm tan\beta$ values.
Table~\ref{sigmavalues} shows the signal significance for different
$\rm m_{H^{+}}$ and $\rm \tnb$ values at $\rm 30~fb^{-1}$. The
extrapolation to large $\rm \tnb$ values shows that a $\rm 5\sigma$
significance for $m_{\rm H^{+}}=170$~GeV/$c^{2}$ is obtained only
for $\rm \tnb \ge 100$.
\begin{table*}[t]
\caption{The signal significance for different $m_{(\rm H^{+})}$ and
$\rm \tnb$ values at $\rm 30~fb^{-1}$ including systematic
uncertainties.} \label{sigmavalues}
\begin{center}
\begin{tabular}{|l|c|c|c|c|c|}
\hline
$ $ & $\rm \tnb = 10 $ & $\rm \tnb = 20 $ & $\rm \tnb = 30 $ & $\rm \tnb = 40 $ & $\rm \tnb = 50 $ \\
\hline
$m_{\rm H^{+}}=140$~GeV/$c^{2}$ & 1.60 & 5.02 & 9.43 & 15.00 & 20.17 \\
\hline
$m_{\rm H^{+}}=150$~GeV/$c^{2}$ & 0.96 & 2.50 & 4.70 & 8.61  & 12.71 \\
\hline
$m_{\rm H^{+}}=160$~GeV/$c^{2}$ & 0.33 & 0.93 & 1.66 & 3.76  & 5.98 \\
\hline
$m_{\rm H^{+}}=170$~GeV/$c^{2}$ & 0.07 & 0.21 & 0.46 & 0.75  & 1.19 \\
\hline
\end{tabular}
\end{center}
\end{table*}

\begin{figure*}[t]
\begin{center}
    \resizebox{6cm}{6cm}{\includegraphics{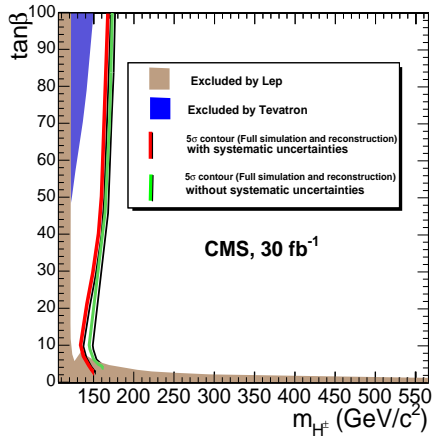}}
  \caption{The $\rm 5 \sigma$ contour in ($m_{\rm H^{+}},\tnb$) plane for the light charged
Higgs discovery including the effect of systematic uncertainties.}
  \label{contourhplus}
\end{center}
\end{figure*}

The results listed in Table~\ref{sigmavalues} are shown in the form
of a $\rm 5 \sigma$ discovery contour in Fig.~\ref{contourhplus}.
The result of introducing the systematic uncertainties in
calculation of the signal significance reduces the observability of
$m_{\rm H^{\pm}}$ for $\rm \tnb<50$. For $\rm \tnb>50$ the signal
cross section increases and the effect of the uncertainties becomes
small.
The light charged Higgs discovery potential of CMS was presented in
the $\rm t \bar{t}\ra H^{\pm}W^{\mp}b \bar{b} \rm ~\ra \ell
\nu_{\ell} \tau \nu_{\tau} b \bar{b}$ ($\tau\rightarrow$hadrons)
channel. Results correspond to an integrated luminosity of $\rm
30~fb^{-1}$ including low uminosity pile--up events. The systematic
uncertainties are evaluated and included. It was shown that the
effect of the systematic uncertainties is a decrease of
5-10~GeV/$c^2$ the observable charged Higgs mass for $\rm \tnb<50$.
Since for $m_{\ch}>160$~GeV/$c^2$ the signal rate becomes small the
$\rm 5 \sigma$ discovery contour shows a small growth for
$m_{\ch}\geq 160$~GeV/$c^2$ and the charged Higgs mass of
170~GeV/$c^2$ is observable at $\rm \tnb \simeq 100$.
\section{Acknowledgement}
I would like to thank the IPM conference organizers especially Dr.
Farzan who provided this opportunity for me to have a talk there and
report my recent results in the CMS collaboration.


\begin{thebibliography}{}
%
%
\bibitem{MSSM}{S.P. Martin, A Supersymmetry Primer, hep-ph/9709356}
\bibitem{lep}{ALEPH, DELPHI, L3 and OPAL Collaborations, The LEP Working Group for Higgs Boson Searches, CERN-PH-EP/2006-001, hep-ex/0602042}
\bibitem{directlep}{ALEPH, DELPHI, L3 and OPAL Collaborations, The LEP Working Group for Higgs Boson Searches, LHWG note 2001-05}
\bibitem{d0}{V.M. Abazov et al., hep-ex/0102039 }
\bibitem{cdf}{B. Abbott et al., Phys. Rev. Lett {\bf 82}(4975)1999 }
\bibitem{mynote}{M. Baarmand, M. Hashemi, A. Nikitenko, \bf CMS NOTE 2006-056}
\bibitem{tilman}{E.L. Berger et al., hep-ph/0312286, \url{http://pheno.physics.wisc.edu/~plehn/charged_higgs/}}
\bibitem {ttbar} {M.L. Mangano et al., CERN Yellow Report, ID: 2000-004, Workshop on Standard Model Physics (and more) at the LHC, hep-ph/0003033}
\bibitem{madevent} { \url{http://madgraph.hep.uiuc.edu}}
\bibitem{cmsim}{\url{http://www-collider.physics.ucla.edu/cms/cmsim/}}
\bibitem{oscar}{\url{http://cmsdoc.cern.ch/oscar/}}
\bibitem{orca}{\url{http://cmsdoc.cern.ch/orca/}}
\bibitem{daq}{The Trigger and Data Acquisition Project, Volume II, Technical Design Report, CERN/LHCC 2002-26, CMS TDR 6.2}
\bibitem{Sasha}{A. Nikitenko et al., \bf CMS NOTE 2000-055 }
\bibitem{type1}{A. Nikitenko et al., \bf CMS NOTE 2001-040 }
\end{thebibliography}
\end{document}